\documentclass[useAMS,usenatbib]{mn2e}
\usepackage{url}
\usepackage{graphicx}
\usepackage{verbatim}
\usepackage[usenames]{color}
\usepackage[flushleft]{threeparttable}
\DeclareGraphicsExtensions{.pdf,.png,.jpg,.mps,.eps,.ps}
\usepackage{amsmath}
\usepackage{natbib}
\usepackage{color}
\usepackage{graphicx}
\usepackage{lscape}
\usepackage{gensymb}

\newcommand\nustar{{\it NuSTAR}}

\newcommand\rxte{{\it RXTE}}
\newcommand\xmm{{\it XMM-Newton}}

\newcommand\kev{{\rm~keV}}

\newcommand\kms{\ifmmode {\rm~km\ s}^{-1} \else ~km s$^{-1}$\fi}
\newcommand\Hunit{\ifmmode {\rm~km\ s}^{-1}\ {\rm Mpc}^{-1}
        \else ~km s$^{-1}$ Mpc$^{-1}$\fi}
\newcommand\ctssec{\ifmmode {\rm~count\ s}^{-1} \else ~count s$^{-1}$\fi}
\newcommand\ergsec{\ifmmode {\rm~erg\ s}^{-1} \else
        ~erg s$^{-1}$\fi}
\newcommand\funit{\ifmmode {\rm~erg\ s}^{-1}\;{\rm cm}^{-2} \else
        ~ergs s$^{-1}$ cm$^{-2}$\fi}
\newcommand\phflux{\ifmmode {\rm~photon\ s}^{-1}\;{\rm cm}^{-2}
        \else   ~photon s$^{-1}$ cm$^{-2}$\fi}
\newcommand\efluxA{\ifmmode {\rm~erg\ s}^{-1}\;{\rm cm}^{-2}\;{\rm
        \AA}^{-1} \else ~erg s$^{-1}$ cm$^{-2}$ \AA$^{-1}$\fi}
\newcommand\efluxHz{\ifmmode {\rm~erg\ s}^{-1}\;{\rm cm}^{-2}\;{\rm
        Hz}^{-1} \else ~erg s$^{-1}$ cm$^{-2}$ Hz$^{-1}$\fi}
\newcommand\cc{\ifmmode {\rm~cm}^{-3} \else cm$^{-3}$\fi}
\newcommand\FWHM{\ifmmode {\rm~FWHM} \else ${\rm~FWHM}$\fi}
\newcommand\Msun{\ifmmode M_{\odot} \else $M_{\odot}$\fi}
\newcommand\Lsun{\ifmmode L_{\odot} \else $L_{\odot}$\fi}

\newcommand\hbeta{\ifmmode {\rm H}\beta \else H$\beta$\fi}
\newcommand\Kalpha{\ifmmode {\rm K}\alpha \else K$\alpha$\fi}
\newcommand\nh{\ifmmode N_{\rm H} \else N$_{\rm H}$\fi}

\usepackage{graphicx}
\usepackage{url}
\usepackage{verbatim}
\usepackage{color}

\title[\nustar{} view of 4U~1636-536]{Evidence of disc reflection in the X-ray spectrum of the neutron star low mass X-ray binary  4U~1636-536}
\author[Mondal et al.]{\parbox[]{6.5in}{{Aditya S. Mondal$^{1}\thanks{E-mail: adityas.mondal@visva-bharati.ac.in}$, B. Raychaudhuri$^{1}$, G. C. Dewangan$^{2}$ }  \\
\small
$^{1}$Department of physics, Visva-Bharati, Santiniketan, West Bengal-731235, India \\
$^{2}$Inter-University Centre for  Astronomy \& Astrophysics (IUCAA), Pune, 411007 India \\
}}

\date{\today}
\begin{document}
\maketitle
\begin{abstract}
We present a broadband spectral analysis of the atoll source 4U~1636-536 observed for $\sim92$ ks with \nustar{}. The source was found to be in a low-luminosity state during this observation with $3-79 \kev{}$ X-ray luminosity of $L_{3-79 keV}=(1.03\pm0.01)\times 10^{37}$ ergs/s, assuming a distance of 6 kpc. We have identified and removed twelve type-I X-ray bursts during this observation to study the persistent emission. The continuum is well described by a thermal Comptonization model {\tt nthcomp} with $\Gamma\sim1.9$, $kT_{e}\sim28 \kev{}$, and $kT_{s}\sim0.9\kev{}$. The \nustar{} data reveal a clear signature of disc reflection, a significantly broad Fe-K emission line (around $5-8\kev{}$), and the corresponding reflection hump (around $15-30\kev{}$). We have modeled the data with two relativistically blurred reflection models. Both families of reflection models prefer truncated inner disc radii prior to the ISCO. We find that the inner disc is truncated with an inner radius of $R_{in}=(3.2-4.7)\;R_{ISCO}$ ($\simeq16-24\,R_{g}\: \text{or}\: 36-54$ km). This inner disc radius suggests that the neutron star magnetic field strength is $B\leq2.0\times10^{9}$ G.      
\end{abstract}
\section{introduction}
A neutron star low mass X-ray binary system (NS LMXB) consists of a neutron star (NS) and a low mass ($\leq1\;M_{\odot}$) companion star. When the NS in an NS LMXB accretes matter from the companion star via Roche-lobe overflow, a geometrically thin, optically thick disc-like structure is formed~\citep{1973A&A....24..337S}. The radiation spectrum from the accretion disc, which is usually accompanied by a hot corona is quasi-thermal in nature and is well known to be multicolor blackbody. The inverse Compton scattering of the thermal disc photon generates a power-law spectrum. Moreover, a hot single-temperature blackbody emission may arise from the boundary layer between the inner accretion disc and the NS surface. Hard X-rays (either a power-law continuum or a blackbody component) can illuminate the accretion disc and produce a reflection spectrum which consists of several emission lines and a broad hump-like shape. \\

The fluorescent Fe K$\alpha$ line is the most prominent emission line due to its large cosmic abundance and high fluorescent yield~\citep{2007ApJ...664L.103B, 2008ApJ...674..415C, 2008ApJ...688.1288P, 2009MNRAS.399L...1R, 2015MNRAS.451L..85D}. In the reflection spectrum, a broad hump-like shape is seen which is created by the high energy photons which tend to Compton scatter back out of the disc~\citep{2001MNRAS.327...10B, 2007MNRAS.381.1697R}. Although the Fe K$\alpha$ line is intrinsically narrow as expected, it becomes broad and asymmetric in the X-ray spectrum of the LMXBs due to the Doppler and the Gravitational shift ~\citep{2000PASP..112.1145F}. A profound study of this line profile is important due to its ability to provide information on the inner accretion flow in the NS LMXBs which in turn provides constraints on the structure of the inner disc and inclination. In an NS binary system, the accretion disc may be truncated by a strong stellar magnetic field or by the boundary layer between the disc and the NS outer surface. The upper limit to the radius of the NS is related to the inner disc radius and may constrain the NS EOS~\citep{2000A&A...360L..35P, 2008ApJ...674..415C, 2011MNRAS.415.3247B}. Moreover, the Fe K$\alpha$ line is also used to find out an upper limit to the strength of the magnetic field related to the NS \citep{2019ApJ...873...99L, 2016MNRAS.461.4049D, 2016ApJ...819L..29K}.\\

4U~1636-536 is an atoll type, bursting LMXB consisting of an NS and an 18th magnitude, 0.4 $M_\odot$ companion star \citep{1990A&A...234..181V}. The source has been studied extensively in the literature. The source has exhibited type-I X-ray bursts, double-peaked X-ray bursts, superbursts, Quasi-Periodic Oscillations (QPOs), and millisecond oscillation during thermonuclear bursts \citep{1997ApJ...479L.141W, 2006ApJ...639.1033G, 2002ApJ...577..337S}. It has an orbital period of $\sim 3.8$ hr \citep{1990A&A...234..181V}. Its location has been estimated to be $6 \pm 0.5$ kpc by \citet{2006ApJ...639.1033G} who also studied its type-I X-ray bursts. \citet{2005MNRAS.361..602S} have shown that the source regularly undergoes state transitions on $\sim 40$ days time intervals. This source is known to show QPOs \citep{1997ApJ...479L.141W, 1996ApJ...469L..17Z, 2007MNRAS.376.1139B} at the domain of kHz. The inclination has been suggested to lie in the range $\sim 30\degree-60\degree$ from optical observations \citep{2006MNRAS.373.1235C}. The source is well known to show burst oscillations at 581 Hz which is remarkably coherent \citep{2002ApJ...577..337S}. This is possibly related to the rotation of the NS. The soft X-ray emission, modulated at the QPO frequency of kHz range is known to have a phase lag behind the hard X-ray emission~\citep{1999ApJ...514L..31K}. This lag could be produced by the reprocessing of hard X-rays in a cooler Comptonizing corona with a size of at most a few kilometers. \\

\citet{2008ApJ...688.1288P} analysed three simultaneous \xmm{} and \rxte{} observation of the source 4U~1636-536, once when the source was in the transitional state and twice when it was in the soft state. In all three spectra, they found clear evidence of a broad, asymmetric iron emission line extending over the energy range $4-9\kev{}$. They reported that the line profile is consistent with the relativistically broadened Fe K$\alpha$ emission line from the accretion disc. They found a lower limit of $64\degr$ on the disk inclination in 4U 1636-536 and reported that it is consistent with the $36\degr-74\degr$ constraint on the orbital inclination by \citet{2006MNRAS.373.1235C}. They reported an upper limit of the inner disc radius ($R_{in}$) which is larger than the ISCO while the source was in the transitional and soft state. \citet{2010ApJ...720..205C} analyzed the spectra of 10 NS LMXBs, including three \xmm{} observation of 4U~1636-536, and found that fitting the spectrum with a self-consistent reflection model resulted in larger values for the inner disc radius than when the spectrum was fitted with the phenomenological {\tt Diskline} model. They reported an inner disc radius of $49\pm24\;GM/c^2\;(\sim8\pm4$ $R_{ISCO}$ where $R_{ISCO}=6\:GM/c^2$ for a spin parameter $a=0$) when the source was in the hard (low flux) state but the values were consistent with the ISCO when measured in the soft state. \citet{2010A&A...522A..96N} analyzed the same spectra as \citet{2008ApJ...688.1288P} and \citet{2010ApJ...720..205C} considering pileup and background effects and suggested a symmetric line profile that could be well fitted with a {\tt GAUSSIAN} model. They interpreted the line width as the result of broadening due to Compton scattering in the surface layers of the ion disc. \citet{2013MNRAS.432.1144S} analyzed six \xmm{} observations of 4U~1636-536 with different phenomenological and reflection models. In all these observations, they found a broad Fe emission line at around $6.5\kev{}$. They also found that Fe line profile exhibits a blue wing extending to high energies as observed by \citet{2008ApJ...688.1288P}. Additionally, \citet{2013MNRAS.432.1144S} measured the inner disc radius to be as large as $26.7\:GM/c^2\; (=4.45\;R_{ISCO}$) in the hard state. They reported that in four observations, the primary source of hard X-rays that reflect off the disc was the NS surface/boundary layer, whereas in the other two observations the reflected spectrum came from the corona. They also reported a high inclination of the source in most cases. \citet{2017ApJ...836..140L} analyzed one \nustar{} observation of 4U~1636-536 in the hard state. They constrained the radius of the inner disc to $1.08\pm0.06$ ISCO for $a=0.3$ through the disc reflection modeling. They found a high inclination of $77-80\degr$ for $a=0.0-0.3$. Recently, \citet{2017MNRAS.468.2256W} analyzed the other three \nustar{} observations of 4U~1636-536 while the source was in different states. In all the spectral states, they found prominent positive broad residuals around $5-10\kev{}$. They applied different models to fit the reflected spectra and found a reasonable inclination of $\sim56\degr$ with the {\tt RELXILLLP} model.  \\

In this work, we present a broadband \nustar{} observation of the source 4U~1636-536. We search for the presence of reflection features and place constraints on the position of the inner disc. In the presence of high quality, pile-up free \nustar{} data and with the correct astrophysical model, X-ray reflection spectroscopy can be quite a powerful tool to probe the accretion geometry. Moreover, the good energy resolution of \nustar{} allows us to identify the presence of Comptonization and of a cut-off in the high energy emission. The paper is structured in the following format: Sec.2 presents the observations and the details of data reduction. Sec.3 discusses spectral analysis and results and Sec. 4 provides the discussion of the results.  \\

\begin{figure*}
\centering
\includegraphics[width=8.0cm, angle=0]{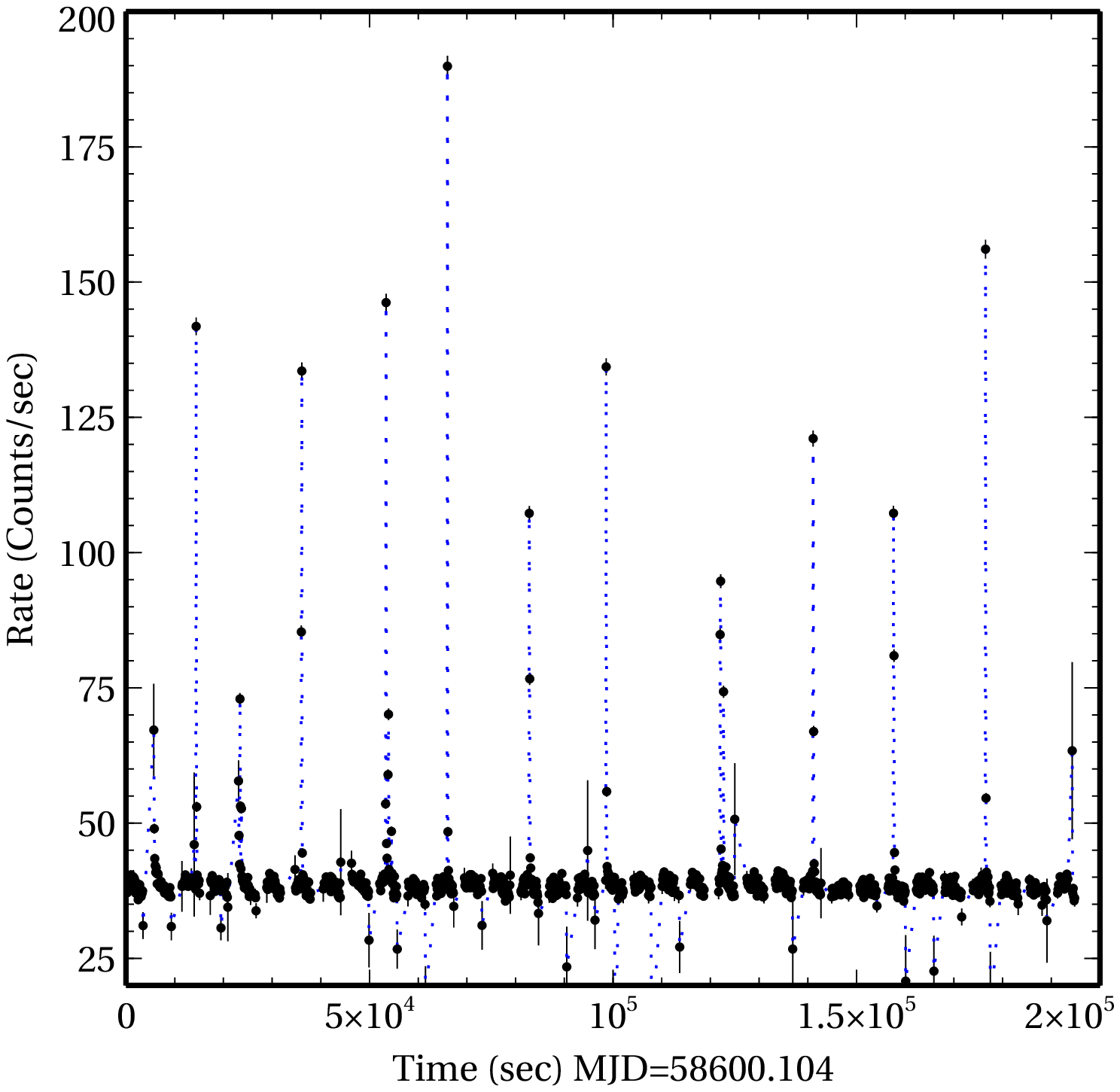}\hspace{1.0cm}
\includegraphics[width=8.0cm, angle=0]{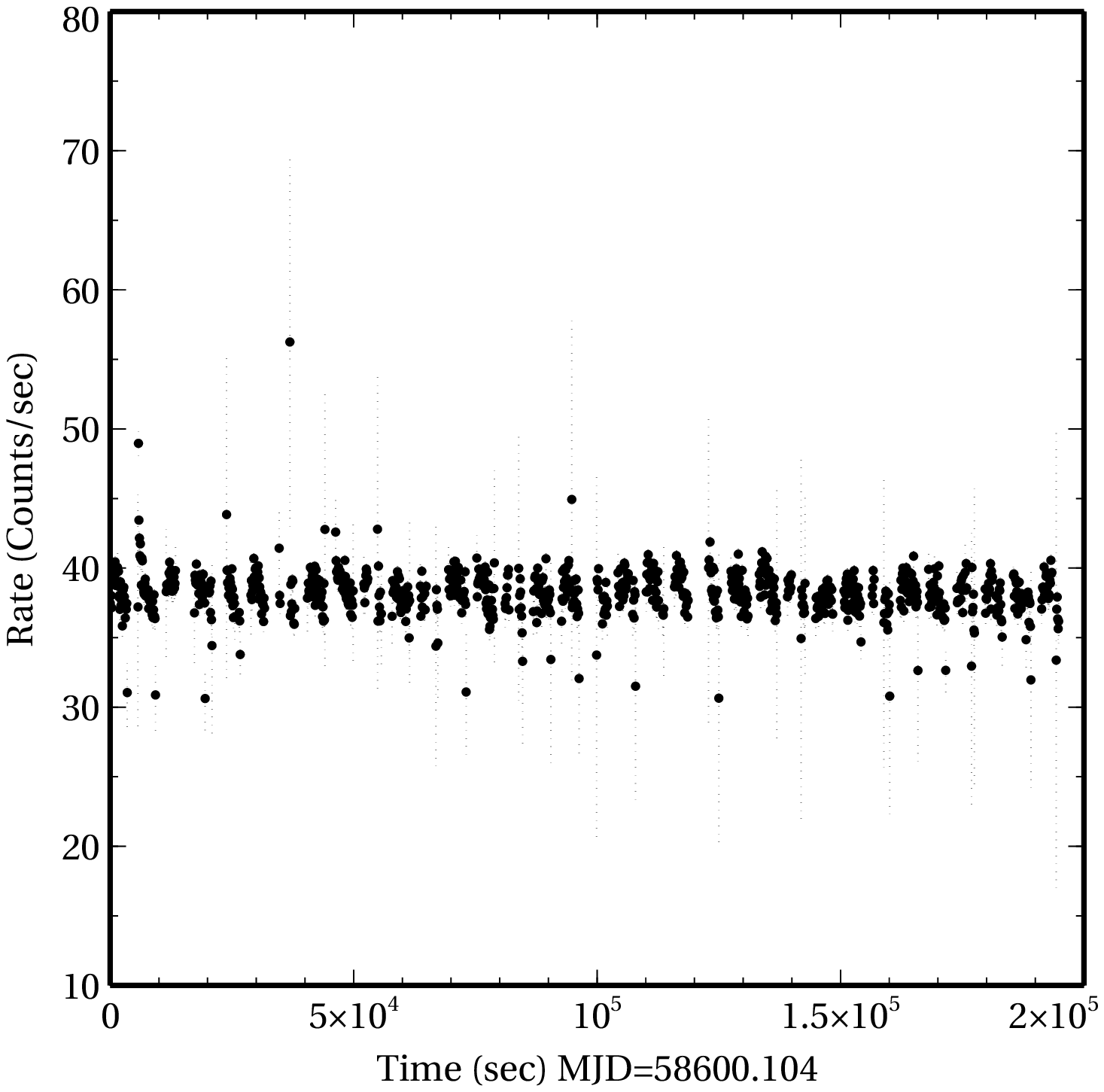}
\caption{Left: $3-79\kev{}$ \nustar{}/FPMA light curve of 4U~1636-536 with a binning of 100 sec. It shows the presence of 12 brief type-I X-ray bursts. Right: Light curve after removing all the type-I X-ray bursts. In this observation, the source detected with an average intensity of $\sim38$ counts/s.} 
\label{Fig1}
\end{figure*}

\section{observation and data reduction}
\nustar{} \citep{2013ApJ...770..103H} observed the source 4U~1636-536 on 2019 April 27 for a total exposure time of $\sim 92$ ks (Obs. ID: $30401014002$). The data were collected with the two co-aligned grazing incidence hard X-ray imaging Focal Plane Modules (FPM) A and B telescopes in the $3-79 \kev$ energy band. \\

The data were reduced with the standard \nustar{} data analysis software ({\tt NuSTARDAS v1.7.1}) and {\tt CALDB} ($v20181030$).
We applied the standard routine {\tt nupipeline} (version v 0.4.6) to filter the event lists. Using the {\tt nuproducts} tool we created lightcurve, spectra, and response files for both the telescopes FPMA and FPMB. To produce a source spectrum for both the telescopes, we extracted a circular region with a radius of 100 arcsec centered around the source position. We extracted the background spectrum from a same-sized radial region away from the source. We grouped the FPMA and the FPMB spectral data with a minimum of 100 counts per bin and fitted the two spectra simultaneously.

\section{spectral fitting}
The source 4U~1636-536 is known to exhibit type-I X-ray bursts. The \nustar{} FPMA and FPMB light curves contain 12 type-I X-ray bursts which are shown in Figure~\ref{Fig1}. We have also shown the light curve after removing the time interval when the type-I X-ray bursts occur. The source was detected at an average intensity of $\sim38$ counts/s during the non-burst period. After excluding all the type-I X-ray bursts, we have fitted both the \nustar{} FPMA and FPMB spectra simultaneously as the initial fits have showed a good agreement between these two spectra. An initial inspection of the FPMA and the FPMB spectra also suggests that the source is detected significantly in the entire energy bandpass of the \nustar{}. We have therefore performed the spectral analysis over the entire $3-79 \kev{}$ energy band using {\tt XSPEC  v 12.9} \citep{1996ASPC..101...17A}. 
Due to flux variations between the detectors, we have added a multiplicative constant in each fit. We have fixed the constant for the FPMA spectrum to unity and allowed it to vary for the FPMB spectrum. A value of $1.002$ has been measured for the FPMB spectrum. We have used the {\tt tbabs} model to account for absorption along the line of sight to the source with the abundance set to {\tt wilm} \citep{2000ApJ...542..914W} and {\tt vern} cross sections \citep{1996ApJ...465..487V}. We have fixed the absorption column density to the \citet{1990ARA&A..28..215D} value of $4.1\times10^{21}$ cm$^{-2}$ as the \nustar{} low-energy bandpass cuts off at $3 \kev{}$ and have found it difficult to constrain from the spectral fits.
All quoted uncertainties in this paper are at $90\%$ of the confidence level if not stated otherwise in particular.

\begin{figure*}
\includegraphics[scale=0.40, angle=-90]{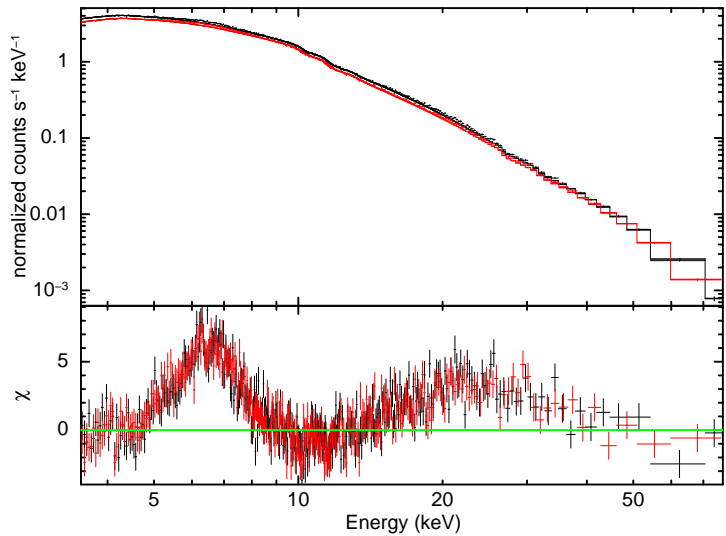}
\caption{\nustar{} (FPMA in black, FPMB in red) unfolded spectra. The data were fit with an absorbed, thermal Comptonization model {\tt nthcomp}. There are prominent residuals at $\sim 5-8$ keV and $\sim 10-20$ keV. Those can be indentified as a broad Fe-K emission line and the corresponding Compton back-scattering hump. The spectral data were rebinned for visual clarity} 
  \label{Fig2}
   \end{figure*} 

\subsection{Continuum modeling}
This \nustar{} observation has detected the source 4U~1636-536 with a luminosity of $\sim1.03\times10^{37}$ erg/s which corresponds to $\sim 5\%$ of the Eddington luminosity ($L_{Edd}$). So, the source is detected in a low-luminosity state with $L/L_{Edd}$ in the range of $0.05-0.1$. The spectra of the low-luminosity state sources are typically characterized by a thermal Comptonization model with an electron temperature ($kT_{e}$) around $25-30\kev{}$ \citep{2000ApJ...533..329B}. In the low-luminosity state, a soft component may be observed probably to represent the unscattered emission from an optically thick accretion disk \citep{2000ApJ...533..329B}.  Additionally, \citet{2007ApJ...667.1073L} also suggested that low/hard state spectra could be modelled with a cutoff power-law component and a single-temperature blackbody component when needed. \\
 
We have modeled the continuum above $3\kev{}$ using only a thermal Comptonization model {\tt nthcomp} \citep{1996MNRAS.283..193Z, 1999MNRAS.309..561Z} which may arise from either a hot corona associated with accretion disc or a boundary layer between the disc and the NS surface (see e.g. \citealt{2016ApJ...819L..29K, 2019ApJ...873...99L}). This thermal Comptonization model has a power-law component with an index $\Gamma$, a low energy cutoff determined by the temperature of the seed-photons ($kT_{s}$) and a high energy roll-over determined by the electron temperature ($kT_{e}$). Here we have assumed that the seed spectrum is a multi-temperature blackbody spectrum emitted from the disc. In our fits, we have allowed both these temperatures and the power-law index to vary. This model, {\tt (tbabs$\times$nthcomp)}, describes the continuum very well but gives a poor fit with $\chi^{2}/dof=4412/1880$. For the {\tt nthcomp} component, we have obtained $\Gamma\sim1.9$, $kT_{e}\sim28\kev{}$, and $kT_{s}\sim0.9\kev{}$. This continuum model has left large positive residuals around $\sim 5-8 \kev{}$ and $\sim 15-30 \kev{}$, which can be identified as a broad Fe-K emission line and the corresponding Compton hump (see Figure~\ref{Fig2}). However, it may be noted that we have not detected any soft blackbody component in the spectrum. It may possibly be due to the inadequate low-energy coverage as in the case of \nustar{}.

\begin{figure*}
   \includegraphics[scale=0.40, angle=-90]{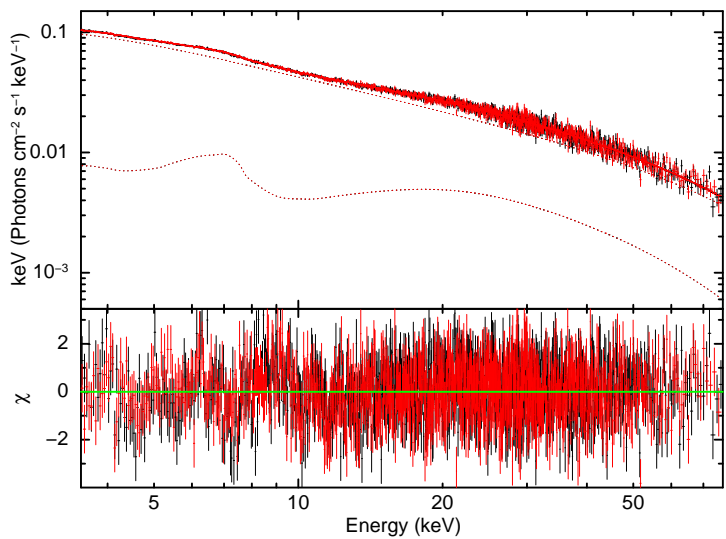}
   \caption{The \nustar{} (FPMA in black, FPMB in red) unfolded spectra of 4U~1636--536 with the best-fitting fitted model consisting of a thermal Comptonization model and a relativistically blurred reflection model (Model 1) i.e.,{\tt TBabs$\times$(nthcomp+highecut*relconv*reflionx)}. Lower panel shows residuals in units of $\sigma$.} 
   \label{Fig3}
   \end{figure*}

\begin{figure}
\includegraphics[width=7.0cm, angle=0]{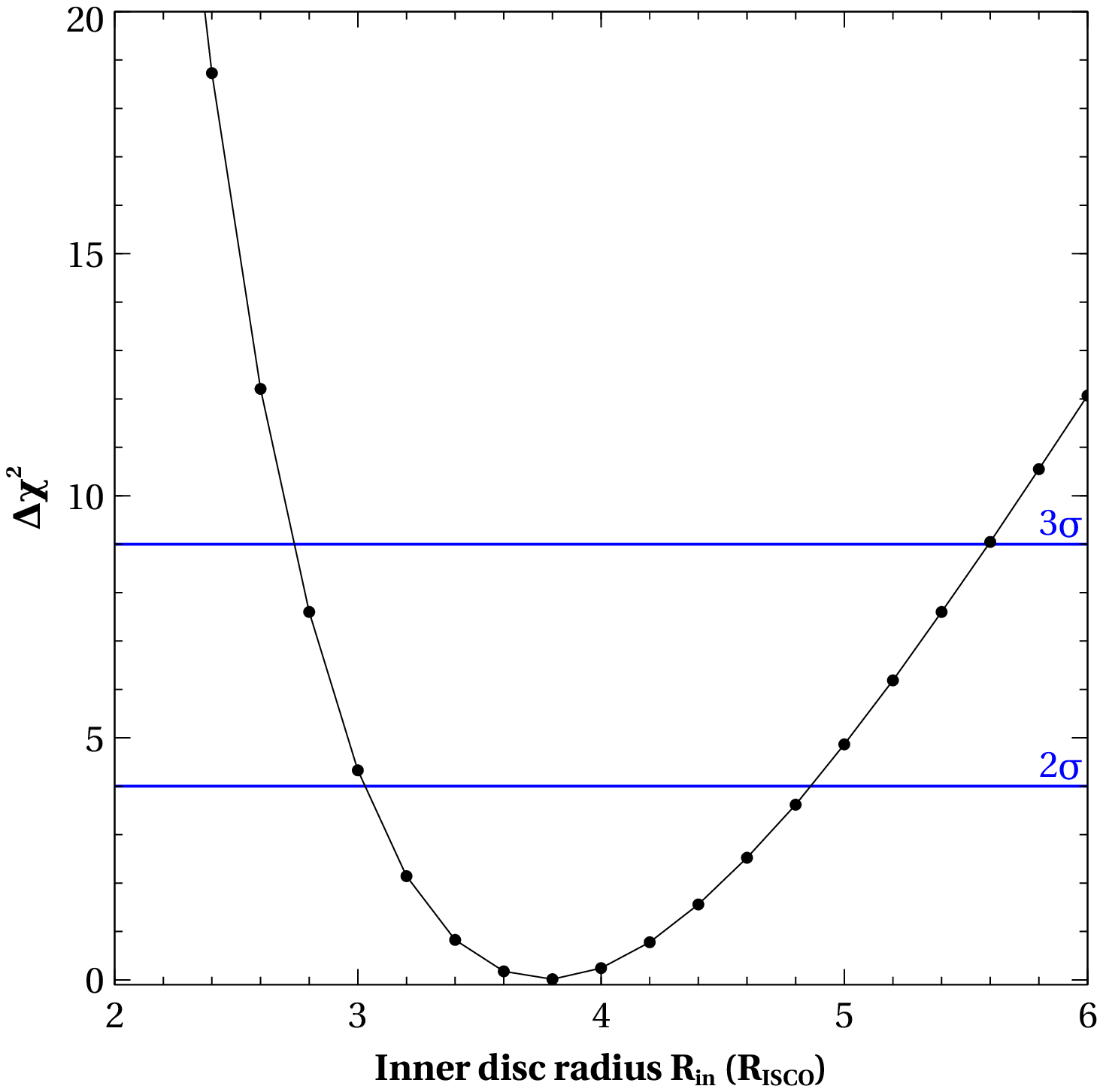}
\caption{Shows the variation of $\Delta\chi^{2}(=\chi^{2}-\chi_{min}^{2})$ as a function of inner disc radius (in the unit of $R_{ISCO}$) obtained from the relativistic reflection model (Model 1). We varied the inner disc radius as a free parameter in between $2\,R_{ISCO}$ to $6\,R_{ISCO}$. The parameter is clearly well constrained by the data. The value of the inner disc radius is inconsistent with the position of the ISCO. Horizontal lines are indicating $2\sigma$ and $3\sigma$ significance level.} 
\label{Fig4}
\end{figure}

\subsection{Reflection Modelling}
As our continuum fit indicates the presence of reflection features (see Figure~\ref{Fig2}), we, therefore, have proceeded by modelling our data with a physical reflection model. We have applied the standard {\tt reflionx} \citep{2005MNRAS.358..211R} model that assumes a high energy exponential cutoff power-law irradiating the accretion disc. The model components of {\tt reflionx} model are as follows: $\Gamma$ is the photon index of the illuminating spectrum, $\xi$ is the disc ionization parameter, $A_{Fe}$ is the iron abundance relative to the solar value, $N_{norm}$ is the normalization of the reflected spectrum and $z$ is the redshift of the source. \\

We have modified this reflection model {\tt reflionx} in such a way that it assumes {\tt nthcomp} illuminating spectrum instead of a cutoff power-law, as our broad-band fits prefer a Comptonized model to describe the continuum spectrum. The cutoff power-law does not have a low-energy cutoff while Comptonization spectra require a low-energy cutoff at the seed photon temperature ($kT_{s}$). Moreover, the high energy cutoff of the illuminating power law in the {\tt reflionx} model is set to $300\kev{}$. Therefore, we have modified {\tt reflionx} in such a manner that it mimics the {\tt nthcomp} continuum. In order to introduce low and high energy cutoff, we have multiplied {\tt reflionx} by a high energy cutoff, {\tt highecut}, with the folding energy $E_{fold}$ set to $\sim 3$ times of the electron temperature $kT_{e}$ and the cutoff energy $E_{cutoff}$ tied to $0.1 \kev{}$. This is how we have introduced a cutoff in the reflection continuum. Moreover, we fixed the photon index ($\Gamma$) of the illuminating spectrum to that of the {\tt nthcomp} component. Thus, we have modified the reflection model {\tt reflionx} in order to reproduce the {\tt nthcomp} continuum by introducing the model component {\tt highecut} (for details see \citealt{2017A&A...600A..24M, 2020MNRAS.494.3177M}). To take relativistic blurring into account, we have convolved {\tt reflionx} with {\tt relconv} component \citep{2010MNRAS.409.1534D}. Now the overall model becomes {\tt TBabs$\times$(nthcomp+highecut*relconv*reflionx)}, hereafter Model 1. The parameters of the {\tt relconv} include the inner and the outer disc emissivity indices ($q_{in}, q_{out}$), break radius ($R_{break}$), the inner and outer disk radii $R_{in}$ and $R_{out}$, the disk inclination ($i$) and the dimensionless spin parameter ($a$). \\

We have imposed a few reasonable conditions when making fits with reflection models. We have assumed an unbroken emissivity profile with a fixed slope of $q=3$, as the slope is not constrained by the data. We have also fixed the outer disc radius $R_{out}$ to $1000\;R_{g}$. We set a redshift of $z=0$ since 4U~1636-536 is a Galactic source. From previous measurements of the NS spin frequency 581 Hz, we have approximated the spin parameter $a=0.27$ as $a\simeq0.47/P_{ms}$ \citep{2000ApJ...531..447B} where $P_{ms}$ is the spin period in ms. Furthermore, we have fixed the disc inclination, $i$, to $60\degr$ as it was poorly constrained when left free to vary (see also \citealt{2008ApJ...688.1288P}). Moreover, prior knowledge of disc inclination can significantly reduce the uncertainty of the measurement of the inner disc radius \citep{2008ApJ...688.1288P}.\\

Adding the relativistic reflection significantly improves our spectral fits with a $\chi^{2}/dof=2178/1876=1.16$. The best fit parameters for the continuum and the reflection spectrum are listed in Table~\ref{parameters}. The reflection component implies a large disc truncation prior to the ISCO at $(3.2-4.7)\;R_{ISCO}$ ($\simeq16-24\,R_{g}\: \text{or}\: 36-54$ km). This is quite consistent with \citet{2010ApJ...720..205C} and \citet{2013MNRAS.432.1144S} when the source was in the hard state. This model yields an intermediate disc ionization of $\xi\sim223$ erg s$^{-1}$ cm which is consistent with $\text{log}\xi\sim (2-3)$ seen in other NS LMXBs (see e.g. \citealt{2010ApJ...720..205C}). The  Fe abundance obtained is consistent with solar composition ($A_{Fe}=1.45\pm0.26$). The fitted spectrum with relativistically blurred reflection model and the residuals are shown in the Figure~\ref{Fig3}. We have computed the distribution of $\Delta\chi^2$ for the parameter inner disc radius ($R_{in}$) using {\tt steppar} command in {\tt xspec} to determine how the goodness-of-fit changed as a function of this parameter. Figure~\ref{Fig4} indicates that $R_{in}$ is well constrained by the data. Moreover, it shows that $R_{in}$ is inconsistent with the position of the ISCO, indicating that it is truncated far from the NS surface.\\

\begin{table*}
   \centering
   \caption{ Best-fitting spectral parameters of the both 2019 and 2015 \nustar{} observations of the source 4U~1636-536 using model: Model 1 {\tt TBabs$\times$(nthcomp+highecut*relconv*reflionx)}.} 
\begin{tabular}{|p{1.6cm}|p{4.2cm}|p{2.5cm}|p{2.5cm}}
    \hline
    Component     & Parameter (unit) &  2019 \nustar{}  &   2015 \nustar{} \\
                  &                  &  \hspace{0.5cm} value  &\hspace{0.5cm} value\\
    \hline
    {\scshape tbabs}    & $N_{H}$($\times 10^{21}\;\text{cm}^{-2}$) &$4.0(f)$ & $4.0(f)$    \\
    {\scshape nthcomp} & $\Gamma$ &  $1.93\pm0.01$ & $2.28\pm0.03$  \\
    & $kT_{e}(\kev)$ &  $28.8_{-2.8}^{+3.9}$ & $18.9_{-2.2}^{+3.1}$ \\
    & $kT_{s}(\kev)$ &  $0.99\pm0.03$ & $1.43_{-0.15}^{+0.23}$ \\
    & norm  & $0.13\pm0.003$  & $0.08\pm0.02$   \\
    {\scshape highecut} & $E_{cut}$($\kev$) & $0.1(f)$ &  $0.1(f)$  \\
    & $E_{fold}$($\kev$)    &  $\simeq3kT_{e}$  & $\simeq3kT_{e}$   \\

   {\scshape relconv} & $i$ (degrees) & $60(f)$ & $86_{-4}^{\dagger}$   \\
   & $R_{in}$($\times R_{ISCO}$) & $3.8_{-0.6}^{+0.9}$ & $1.13\pm0.05$\\
   {\scshape reflionx} & $\xi$(erg cm s$^{-1}$) &  $223_{-6}^{+8}$ & $2338_{-965}^{+1369}$  \\
   & $\Gamma$  & $1.93\pm0.01 $ & $2.28\pm0.03$\\
   & $A_{Fe}$ ($\times \;\text{solar})$   & $1.45\pm0.25$ & $0.37_{-0.09}^{+0.05}$  \\
   & norm ($\times 10^{-5}$)   &  $1.65\pm0.21$ & $1.21_{-0.06}^{+0.14}$\\
   & $F^{*}_{total}$ ($\times 10^{-9}$ ergs/s/cm$^2$) & $2.4\pm 0.01$ & $1.9\pm0.02$\\
   & $F_{nthcomp}$ ($\times 10^{-9}$ ergs/s/cm$^2$)&  $1.9 \pm 0.01$ & $1.2\pm 0.01$\\
   & $F_{reflionx}$ ($\times 10^{-9}$ ergs/s/cm$^2$) &  $0.5 \pm 0.01$ & $0.7\pm0.01$ \\
   & $L_{3-79 keV}$ ($\times 10^{37}$ ergs/s) & $1.03 \pm 0.01$ & $0.82\pm0.02$ \\	
  
   \hline 
    & $\chi^{2}/dof$ & $2178/1876$ & $998/1040$  \\
    \hline
  \end{tabular}\label{parameters} \\
{\bf Note:} Here we have used the standard {\tt reflionx} model that assumes a high energy exponential cutoff power-law irradiating the accretion disc, modified in such a manner that it  mimics the {\tt nthcomp} continuum (see text).   
The outer radius of the {\tt relconv} spectral component was fixed to $1000\;R_{g}$. We fixed emissivity index $q=3$. The $E_{fold}$ parameter is fitted to be 3 times the $kT_{e}$. $\dagger$ indicates that the parameter is at the hard-coded limit of the model.\\
$^{*}$All the unabsorbed fluxes are calculated in the energy band $3-79\kev$
\end{table*}

It may be noted that the reflection features are better explained by the utilization of the self-consistent reflection model {\tt RELXILL}. Some new flavors of the {\tt RELXILL} model are available today. A flavor of the {\tt RELXILL} model, {\tt RELXILLCP}, uses {\tt nthcomp} as a illuminating continuum. In this model, the reflection spectrum is self-consistently calculated using the more physical illumination continuum calculated with {\tt nthcomp}. The downside of {\tt RELXILLCP} is the fixed seed photon temperature of $0.05$\kev{}. We detected a higher seed photon temperature of $\sim 0.9$ keV in our continuum fit with {\tt nthcomp}. Still, we attempted to fit the spectrum with {\tt RELXILLCP} model. The overall model now becomes {\tt TBabs$\times$(nthcomp+RELXILLCP)}, hereafter Model 2. It would be helpful to check how the best-fit parameter values depend on the choice of the model since neither of them is ideal. The parameters of {\tt RELXILLCP} are as follows: the inner and the outer disc emissivity indices ($q_{in}, q_{out}$), break radius ($R_{break}$), the inner and outer disk radii $R_{in}$ and $R_{out}$, respectively, the disk inclination ($i$), the dimensionless spin parameter ($a$), the redshift to the source ($z$), which we fixed to $0$, the photon index of the power-law ($\Gamma$), the ionization parameter (log$\xi$), the iron abundance ($A_{Fe}$), the coronal electron temperature ($kT_{e}$), the reflection fraction ($f_{refl}$), and the normalization. For this fit, the emissivity profile is assumed to follow a power law with an index fixed at the canonical value of $q=3$. We have tied the emissivity indices to create one emissivity index and fixed the outer disc radius $R_{out}$ to $ 1000\;R_{g}$. We have fixed the spin parameter $a=0.27$. To get a consistent fit, we tied the {\tt RELXILLCP} photon index $\Gamma$ and electron temperature $kT_{e}$ to that of the {\tt nthcomp} component. \\

\begin{figure*}
   \includegraphics[scale=0.40, angle=-90]{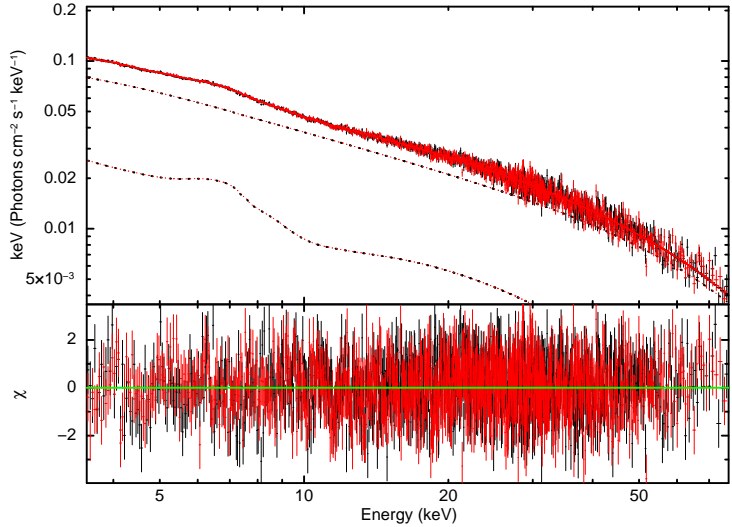}
   \caption{The \nustar{} (FPMA in black, FPMB in red) unfolded spectra of 4U~1636--536 with the best-fitting fitted model consisting of a thermal Comptonization model and a self-consistent reflection model (Model 2) i.e.,{\tt TBabs$\times$(nthcomp+RELXILLCP)}. Lower panel shows residuals in units of $\sigma$.} 
   \label{Fig5}
   \end{figure*}

\begin{figure*}
\includegraphics[width=6.0cm, angle=0]{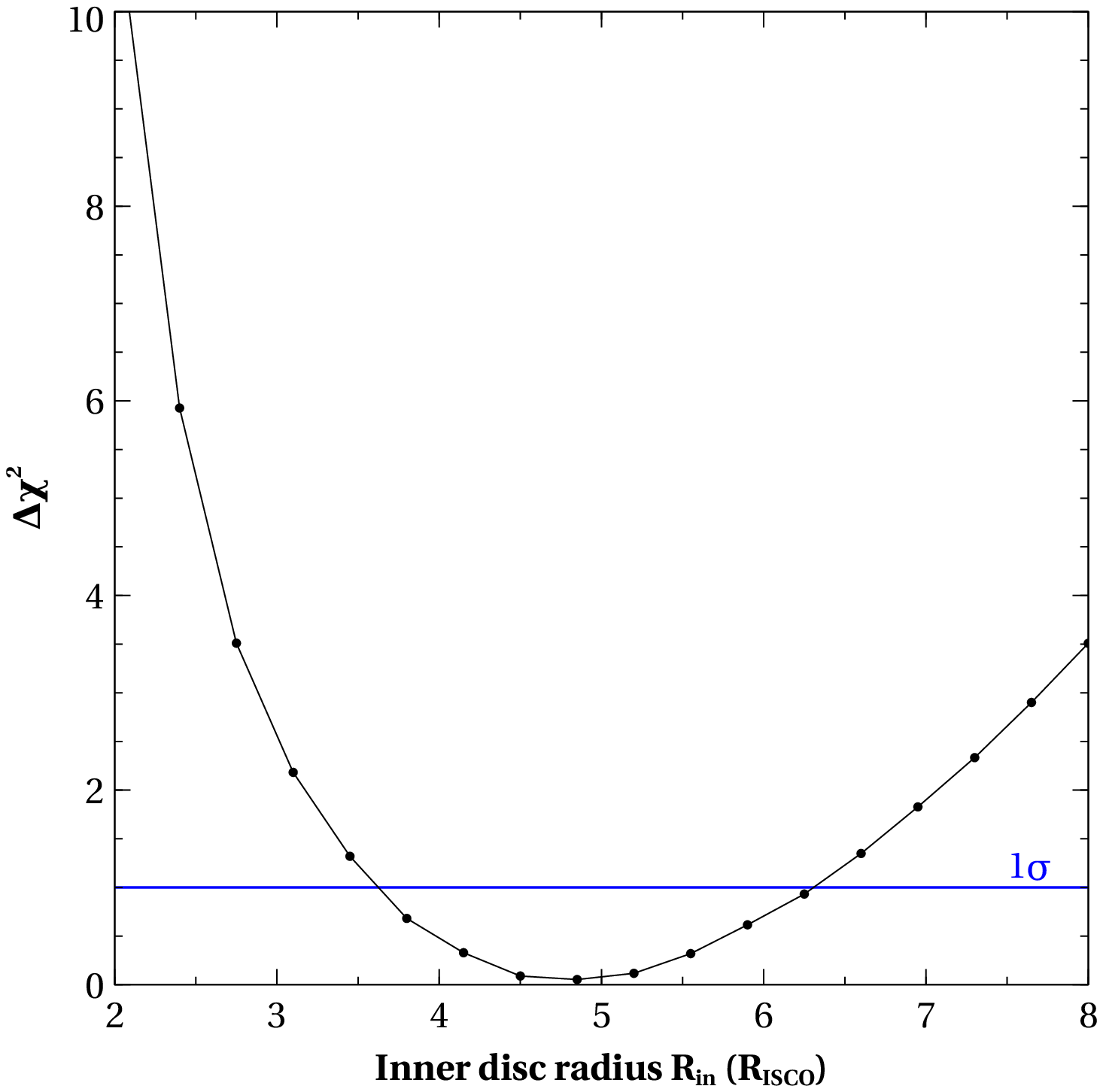}\hspace{2cm}
\includegraphics[width=6.0cm, angle=0]{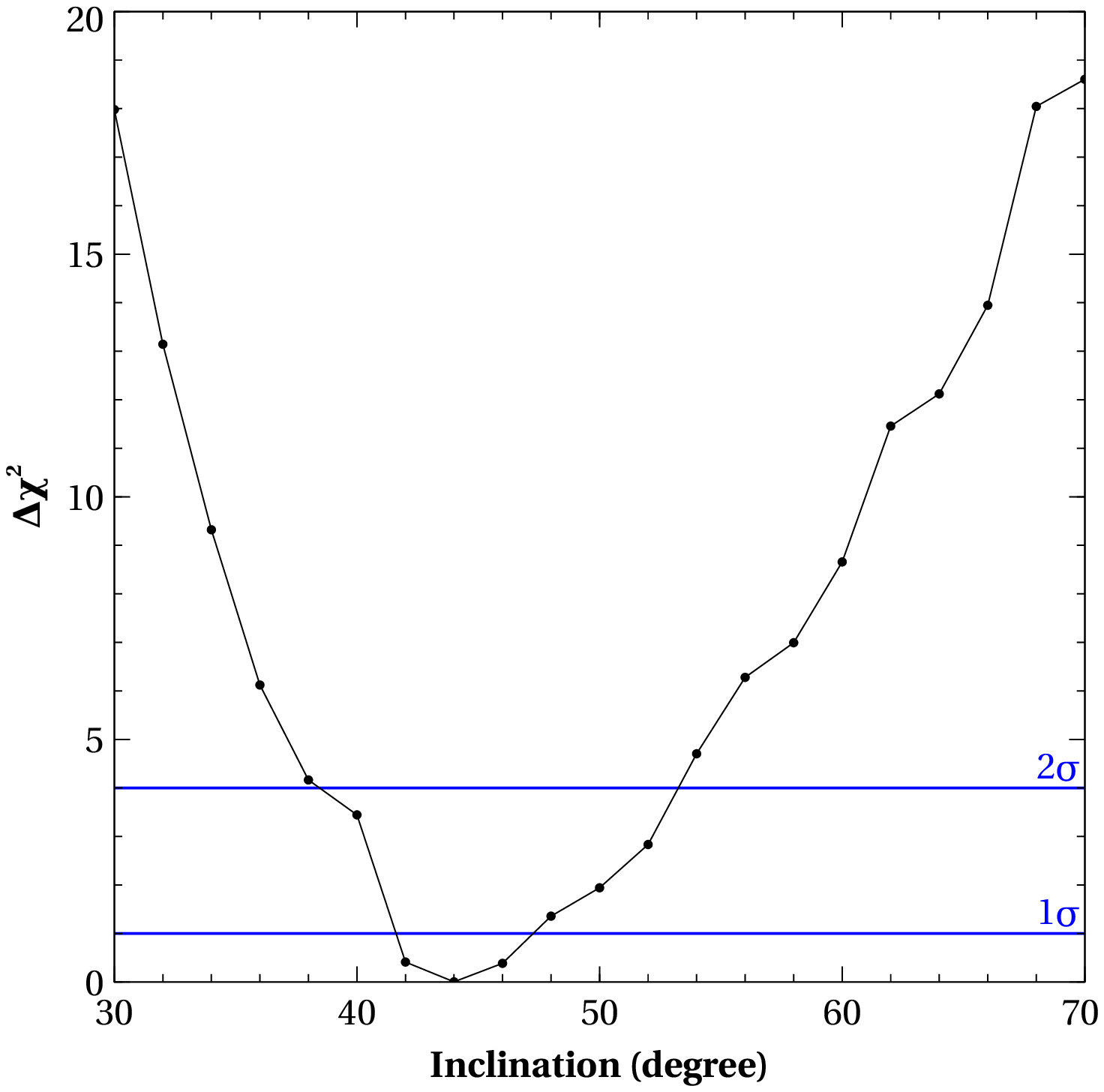}
\caption{Shows the variation of $\Delta\chi^{2}(=\chi^{2}-\chi_{min}^{2})$ as a function of inner disc radius (in the unit of $R_{ISCO}$) and inclination obtained from the self-consistent reflection model {\tt RELXILLCP} in Model 2. Both the parameter are well constrained by the data. Horizontal lines are indicating $1\sigma$ and $2\sigma$ significance level.}
\label{Fig6}
\end{figure*}

The addition of this model improved the overall fit significantly to $\chi^{2}/dof=2108/1874=1.13$. Most of the parameters of Model 2 follow the same trend as those of Model 1. The inner disc radius $R_{in}=(2.1 - 6.7)\;R_{ISCO}$ is consistent with the value obtained with the previous modeling of Model 1. This is also consistent with \citet{2010ApJ...720..205C} and \citet{2013MNRAS.432.1144S} when the source was in the hard state. This demonstrates that the measurement of the inner disc radius is robust. Moreover, the photon index $\Gamma$, iron abundance $A_{Fe}$,  and the electron temperature $kT_{e}$ values are also consistent with the values of the parameters of Model 1.  Importantly, Model 2 gives a reasonable disc inclination of $\sim 44\degr$ which is consistent with \citet{2006MNRAS.373.1235C} and \citet{2017MNRAS.468.2256W} who reported an inclination of $\sim56\degr$ with the {\tt RELXILLLP} model. Also, different authors reported it larger than $64\degr$ in all cases (\citealt{2008ApJ...688.1288P, 2013MNRAS.432.1144S, 2017ApJ...836..140L}). However, with this model, we found a higher disc ionization parameter (log$\xi$) compared to that in Model 1 but comparable with other similar kinds of accreting sources. At the same time, We are unable to constrain the lower limit of the reflection fraction $f_{refl}$ value. The best-fit parameter values are given in Table~\ref{parameters1}. The best-fit spectrum, individual components, and residuals are shown in Figure~\ref{Fig5}. Moreover, the distribution of $\Delta\chi^2$ is calculated using the command {\tt steppar} in {\tt XSPEC} over 20 steps in the inner disc radius ($R_{in}$) as well as in the inclination angle ($i$). The left and right panel of Figure~\ref{Fig6} shows the $\Delta\chi^2$ of the fit versus the $R_{in}$ and $i$, respectively for Model 2.

\begin{figure*}
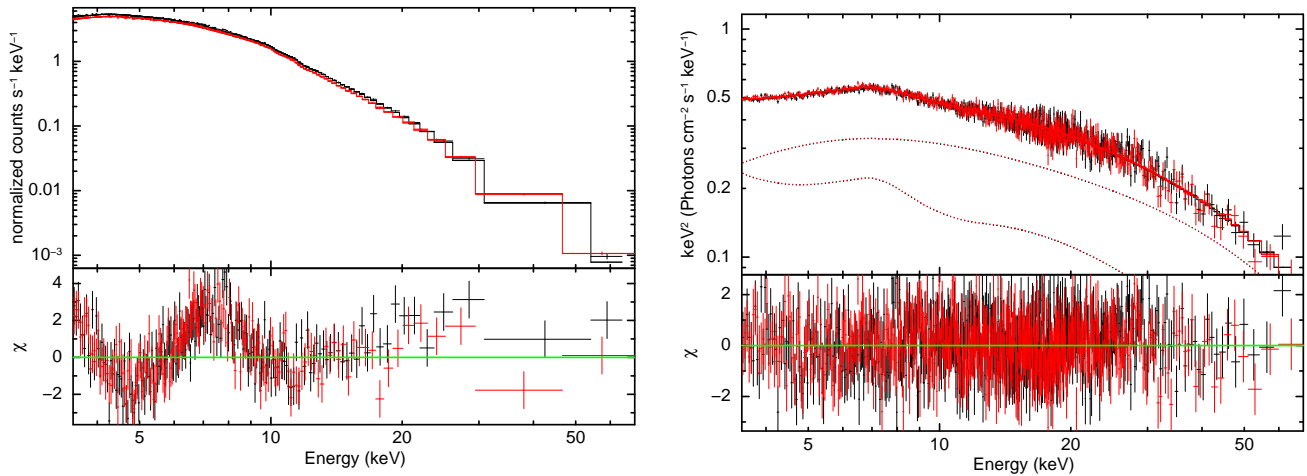

   \includegraphics[scale=0.35, angle=-90]{fig8.ps}
   \includegraphics[scale=0.35, angle=-90]{fig9.ps}
   \caption{Left: The 2015 \nustar{} observation shows the strong reflection features when continuum is fitted with {\tt TBabs$\times$ nthcomp}. Right: the \nustar{} (FPMA in black, FPMB in red) unfolded spectra of 4U~1636--536 with the previous best-fitting model (Model 1). Lower panel shows residuals in units of $\sigma$. In both the cases, the spectra have been rebinned for plotting purposes.}
   \label{Fig7}
   \end{figure*}

\section{2015 \nustar{} observation}
Apart from the recent 2019 \nustar{} observation stated in the last section, it observed the source on 2015 June 06 (Obs ID: 30101024002) for a total exposure of $\sim 20$ ks with a net count rate of $34$ counts/s. This coordinated \nustar{} observation was previously analyzed by \citet{2017ApJ...836..140L}. Here, we have re-analyzed the 2015 \nustar{} observation in the very similar approach as done above for the recent 2019 \nustar{} observation to see if there is any change between the observations. In this connection, it may be noted that the other three \nustar{} observations (between 2015 August 25 and September 18) which are subsequent to the observation in \citet{2017ApJ...836..140L} was analyzed by \citet{2017MNRAS.468.2256W}. In between these three observations, the source evolves from the soft to the transitional, and the hard state \citep{2017MNRAS.468.2256W}. In this observation, the source was in the hard state as reported by \citet{2017ApJ...836..140L}. \\

\subsection{Spectral fitting}
\begin{figure}
\includegraphics[width=7.0cm, angle=0]{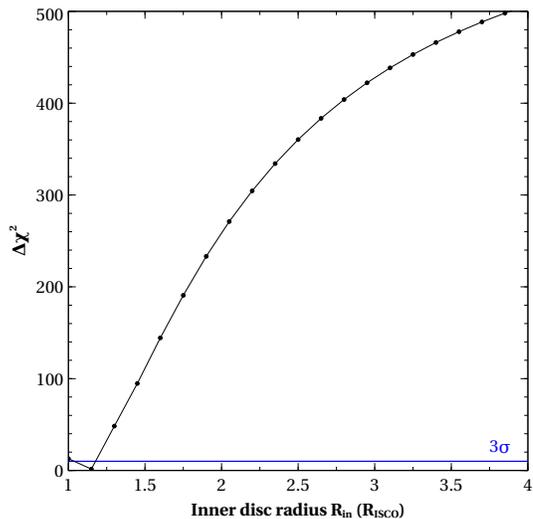}
\caption{The 2015 \nustar{} data: This plot shows the variation of $\Delta\chi^{2}(=\chi^{2}-\chi_{min}^{2})$ as a function of inner disc radius (in the unit of $R_{ISCO}$) obtained from the relativistic reflection model (Model 1). We varied the inner disc radius as a free parameter in between $1\,R_{ISCO}$ to $4\,R_{ISCO}$. The parameter is clearly well constrained by the data. The value of the inner disc radius is consistent with the position of the ISCO. Horizontal line is indicating $3\sigma$ significance level.} 
\label{Fig8}
\end{figure}

We processed the 2015 \nustar{} data in the similar fashion as described before for the 2019 \nustar{} observation. There were six type I X-ray bursts in this observation. We created good time intervals to eliminate the bursts from the spectrum of the persistent emission. We tried to fit the continuum spectrum from 3 to 70 $\kev{}$ with the absorbed thermal Comptonization component {\tt nthcomp} assuming that the seed spectrum is a multi-temperature blackbody spectrum emitted from the disc. This model, {\tt tbabs$\times$nthcomp}, describes the continuum well but left large positive residuals around $5-10\kev{}$ and comparatively less positive residuals around $20-30\kev{}$ (see left panel of Figure~\ref{Fig7}) . This time the reflection hump at high energies is not as clear as detected for the 2019 \nustar{} observation. The other three \nustar{} observation, analyzed by \citet{2017MNRAS.468.2256W}, also did not show the presence of a clear reflection hump. For this observation, the features can be seen in the left panel of Figure~\ref{Fig7}. To describe the reflection features and relativistic effects properly, we employed our Model 1 (described earlier) to the data with very similar conditions imposed on some parameters. This model describes the spectrum successfully with $\chi^{2}/dof=998/1040=0.96$. It predicts a small inner disc radius of $R_{in}=(1.08 - 1.18)\;R_{ISCO}$ which is consistent with \citet{2017ApJ...836..140L}. This fit returns a higher disc inclination of $\sim86\degr$ also consistent with \citet{2017ApJ...836..140L}. The ionization parameter is comparatively high but comparable with other similar kinds of accreting sources. The iron abundance is less than 1. The best-fitting parameters of Model 1 for this observation are presented in Table~\ref{parameters}. The best-fit spectrum with individual components can be seen in the right panel of Figure~\ref{Fig7}. The variation of $\Delta\chi^{2}$ of the fit versus the inner disc radius $R_{in}$ is shown in Figure~\ref{Fig8}.

\begin{table*}
   \centering
\caption{Best-fitting spectral parameters of the 2019 \nustar{} observation of the source 4U~1636-536 using model: Model 2 {\tt TBabs$\times$(nthcomp+RELXILLCP)}.} 
\begin{tabular}{|p{1.6cm}|p{4.2cm}|p{1.7cm}|}
    \hline
    Component     & Parameter (unit) & Value \\
    \hline
    {\scshape tbabs}    & $N_{H}$($\times 10^{21}\;\text{cm}^{-2}$) &$4.0(f)$     \\
    {\scshape nthcomp} & $\Gamma$ &  $1.83\pm0.01$   \\
    & $kT_{e}(\kev)$ &  $22.3_{-1.1}^{+1.5}$ \\
    & $kT_{s}(\kev)$ &  $0.89_{-0.04}^{+0.12}$  \\
    & norm  & $0.11\pm0.01$     \\
    {\scshape relxillcp} & $i$ (degrees) & $44_{-6}^{+8}$    \\
    & $R_{in}$($\times R_{ISCO}$) & $4.8_{-2.7}^{+1.9}$\\
    & $log\xi$(erg cm s$^{-1}$) &  $3.58\pm0.12$  \\
    & $\Gamma$  & $1.83\pm0.01 $  \\
    & $A_{Fe}$ ($\times \;\text{solar})$   & $2.06\pm0.50$  \\
    & $kT_{e}(\kev)$ &  $22.3_{-1.1}^{+1.5}$ \\
    & $f_{refl}$   & $2.57_{\dagger}^{+1.1}$ \\
    & norm ($\times 10^{-4}$)   &  $1.44_{-0.81}^{+1.71}$ \\
   
   \hline 
    & $\chi^{2}/dof$ & $2108/1874$   \\
    \hline
  \end{tabular}\label{parameters1} \\
{\bf Note:} Here we have used a flavor of the {\tt RELXILL} model, {\tt RELXILLCP}, uses {\tt nthcomp} as a illuminating continuum. The outer radius of the {\tt RELXILLCP} spectral component was fixed to $1000\;R_{g}$. We fixed emissivity index $q=3$. $\dagger$ indicates that the lower bound of the parameter is not well constrained by the data. \\

\end{table*}

\section{Discussion}
We have presented here a broadband spectral analysis of the \nustar{} observation of the source 4U~1636-536, aimed to study the reflection spectrum and to constraint the accretion geometry. The continuum spectrum is hard and well described by a thermal Comptonization model {\tt nthcomp} with $\Gamma\sim1.9$, $kT_{e}\sim28 \kev{}$, and $kT_{s}\sim0.9\kev{}$. For the first time, we have detected a clear signature of strong disc reflection in its spectrum. The most prominent disc reflection features, a broad Fe-K emission line around $5-8 \kev{}$ and the corresponding Compton hump around $15-30 \kev{}$ are clearly visible in the spectrum. It must be noted that, previous \nustar{} observations analyzed by \citet{2017MNRAS.468.2256W} and \citet{2017ApJ...836..140L} did not show the presence of a clear Compton hump in the X-ray spectra. A correct choice of self-consistent relativistically blurred disc reflection model helps us to determine the position of the inner disc along with some other important NS parameters. The $3-79 \kev{}$ persistent spectrum is well described by a combination of the Comptonization model {\tt nthcomp} and a relativistic reflection of this Comptonized emission. We have modelled the data with two relativistically blurred reflection models, {\tt relconv*reflionx} and {\tt RELXILLCP}. Both the models assume that a thermal Comptonization spectrum is irradiating the disc. It may be physically interpreted as the region of main energy release, where hard X-rays are produced would be either an optically thin boundary layer between the disc and the NS surface or a hot corona associated with the disc. A part of this hard X-ray emission may illuminate the accretion disc and produce the reflection spectrum.   \\

The source was detected at an average intensity of $\sim38$ counts/s during the non-burst state. Our best-fit model yield an unabsorbed flux in the $3-79\kev{}$ band of $F_{3-79}\sim2.4\times 10^{-9}$ ergs s$^{-1}$ cm$^{-2}$.
The source was observed in a low luminosity and hard spectral state (low/hard) during this observation and we measured a $3-79\kev{}$ luminosity of $L_{X}\sim1.03\times 10^{37}$ ergs s$^{-1}$ which corresponds to $\sim 5\%$ of the Eddington luminosity assuming a distance of 6 kpc. It suggests that the rate of any outflow in 4U~1636-536 is significantly below the Eddington mass accretion rate. It also allows us to study the disc reflection of NS LMXBs in a relatively low accretion regime. \\

From the reflection spectrum (with Model 1), we have measured a inner disc radius of $R_{in}=(3.2-4.7)\:R_{ISCO}$, given that $R_{ISCO}=5.1\:GM/c^2$ for an NS spinning at $a\simeq0.3$. This would correspond to $R_{in}=(16-24)\,R_{g}$ or $(36-54)$ km for a $1.5M_{\odot}$ NS. Another self-consistent reflection model (our Model 2) prefers an inner disc radius of $R_{in}=(2.1 - 6.7)\:R_{ISCO}$ which is consistent with the value obtained by using Model 1. So, both families of reflection models prefer an inner disc radius larger than ISCO. Moreover, the values are consistent with \citet{2010ApJ...720..205C} and \citet{2013MNRAS.432.1144S} when the source was in the hard spectral state. It indicates that the disc is truncated at a large distance away from the NS surface. The disc has a relatively low ionization ($\xi\sim223$ erg s$^{-1}$ cm) and iron abundance is comparable to the solar abundance ($A_{Fe}\sim1.4$). However, the {\tt RELXILLCP} model in Model 2 predicts a higher disc ionization of log$\xi\sim3.5$ erg s$^{-1}$ cm. In our modeling of the reflection spectrum (using Model 1), we fixed the inclination angle to $60\degr$ as it prefers a value close to $90\degr$ when we left it free. A very high inclination ($\sim90\degr$) of the accretion disc is at odds with the fact that no dips or eclipses have been observed in this source. Previously, high inclination angles have been reported by different authors (\citealt{2008ApJ...688.1288P, 2013MNRAS.432.1144S, 2017ApJ...836..140L}). In our case, the reflection modeling with {\tt RELXILLCP} in Model 2 gives a reasonable inclination of $\sim44\degr$ which is quite similar ($\sim 56\degr$) as reported by \citet{2017MNRAS.468.2256W} using the {\tt RELXILLLP} model. During this observation, the source was in the hard state and the corresponding electron temperature ($kT_{e}$) of the corona was $\sim28\kev{}$. This amount of electron temperature indicates that the corresponding cutoff energy ($E_{cut}$) is also very high which is $\simeq(2-3)\;kT_{e}$. The high value of $E_{cut}$ indicates that the illuminating source that produces the reflection component is essentially the corona. \\

We have detected a significantly broader Fe K$\alpha$ emission line ($\sigma\sim0.98\kev{}$) from this source. The broadness of the line requires it to be located deep within the potential well where the orbital velocities are mildly relativistic (see also \citealt{2016ApJ...819L..29K}). Moreover, the observed symmetric nature (a broad blue wing in addition to the broad red wing) of the Fe K$\alpha$ line profile requires it to be located far enough away as to not suffer severe relativistic Doppler beaming (see also \citealt{2016ApJ...819L..29K}). \citet{2008ApJ...688.1288P} suggests that such a broader Fe K$\alpha$ line needs a higher disc inclination of around $60\degr-70\degr$. \\

To place this recent 2019 \nustar{} observation in context, we have re-analyzed another \nustar{} observation (carried out in 2015) which was previously analyzed by \citet{2017ApJ...836..140L}. Like the recent 2019 \nustar{} observation, this particular observation was also in the hard spectral state as reported by \citet{2017ApJ...836..140L}. In this analysis, we have adopted the same models and approach as previously done to investigate if there is any change between the observations. This also allows us to probe how the choice of reflection and continuum models can affect the conclusions about the disk parameters inferred from the reflection component. We found that our results (obtained with Model 1) are more in line with those already found by \citet{2017ApJ...836..140L}. In particular, we found an inner disc radius to be $\sim1.13\;R_{ISCO}$ with an inclination angle $\sim85\degr$. Additionally, the photon index, high energy cut-off, and ionization parameters are largely consistent with the values found by \citet{2017ApJ...836..140L} . However, \citet{2017ApJ...836..140L} reported an iron overabundance of $4.5-5.0$ relative to solar value. According to them, it is unlikely that the abundance of iron is that high in the low mass companion star in 4U~1636-536. Our model estimates an iron abundance of $\sim 0.5$ relative to the solar value which is quite reasonable for the low mass companion star. Further, we found that the major difference between these two observations lies in the fact that the recent 2019 \nustar{} observation predicts a truncated inner disc radius,  whereas 2015 \nustar{} observation predicts a small inner disc radius consistent with ISCO. The standard accretion disc model suggests that the spectral state may be associated with the level of disc truncation \citep{2007A&ARv..15....1D}. Here, both the \nustar{} observations under consideration were in the hard spectral state but a different level of disc truncation was observed. This behavior is quite puzzling and further critical analysis in this direction is needed to find how similar kinds of spectral states give rise to the different levels of disc truncation. \\

Disc truncation is likely the result of the presence of a boundary layer that lies between the disc and the NS surface or the associated magnetic field of the NS. Here we estimate the mass accretion rate ($\dot{m}$) per unit area at the NS surface using Equation (2) of \citet{2008ApJS..179..360G}. The estimated value of $\dot{m}$ during this observation to be $\sim1.6\times10^{-9}\;M_{\odot}\;\text{y}^{-1}$ using the persistent flux $F_{p}=2.4\times 10^{-9}$ erg~s$^{-1}$ cm$^{-2}$ and assuming the bolometric correction $c_{bol}$ is $\sim 1.38$ for the nonpulsing sources \citep{2008ApJS..179..360G} and considering $1+z=1.31$ for a NS with mass 1.5 $M_{\odot}$ and radius 10 km where $z$ is the surface redshift. At this mass accretion rate, we estimate the maximum radial extent ($R_\text{max}$) of the boundary layer region using Equation (2) of \citet{2001ApJ...547..355P}. It estimates a maximum radial extent of $\sim 5.4\:R_{g}$ for the boundary layer (assuming $M_{NS}=1.5\:M_{\odot}$ and $R_{NS}=10$ km). The extent of the boundary layer region is small to account for the disc position. It may be because it does not account for a spin and viscous effects in this layer. \\

The inferred radial extension of the boundary layer is small compared to the disc truncation radius and the magnetic field of the NS would be responsible for the disc truncation. We can use our measured inner disc radius to estimate an upper limit for the magnetic field strength of the NS. We use Equation (1) of \citet{2009ApJ...694L..21C} to calculate the magnetic dipole moment ($\mu$). We estimate a bolometric flux of $F_{bol}\simeq2.94\times10^{-9}$ erg~cm$^{-2}$ s$^{-1}$ by extrapolating the best fit over the $0.01-100 \kev{}$ range. We assume an NS of mass $1.5\:M_{\odot}$, radius $10$ km, and distance of $6$ kpc. We keep similar assumptions regarding the geometrical and efficiency parameters as in \citet{2009ApJ...694L..21C}: $k_{A}=1$ which is a factor depending on the geometry, spherical or disk-like, of the accretion flow, $f_{ang}=1$ which is known as the anisotropy correction factor and accretion efficiency in the Schwarzschild metric $\eta=0.1$. The constraint that $R_{in}\leq24\;R_{g}$ from the best fit model, then yields $B\leq2.0\times10^{9}$ G at the magnetic poles. \\

In this observation, we have observed 12 brief ($10-100$s) X-ray bursts. Accretion onto NS in LMXBs provides the fuel for thermonuclear burning that powers X-ray bursts. Therefore, the detection of X-ay bursts indicates that accreted material is still reaching the surface of 4U~1636-536, even when the disc is truncated at a larger distance. This behavior can be explained with the model proposed by \citet{1991ApJ...372L..87K}. They suggest that in this scenario accreted material can free-fall crossing the 'gap' between the disc and the NS surface and then strike the NS surface, creating a hot accretion belt with a temperature inversion. According to \citet{1977ApJ...217..197L} the accreted material can reach the NS via a magnetic gate but it requires a relatively low magnetic field. If the accreted material is channeled along the magnetic field lines, it would cause a hot spot on the magnetic pole \citep{2009ApJ...706..417L}. However, any further possible mechanisms for accretion onto the NS are out of the scope of this work. Moreover, these studies can further be extended to investigate how the accretion disc responds to an X-ray burst and the impact of bursts on the accretion disc dynamics.

\section{Data availability}
This data set with Obs. ID: $30401014002$ dated 27.04.2019 is in public domain put by NASA at their website https://heasarc.gsfc.nasa.gov. The public date is 08.05.2020. 

\section{Acknowledgements}
We thank the referee for many invaluable comments and suggestions. This research has made use of data and/or software provided by the High Energy Astrophysics Science Archive Research Centre (HEASARC). This research also has made use of the \nustar{} data analysis software ({\tt NuSTARDAS}) jointly developed by the ASI science center (ASDC, Italy) and the California Institute of Technology (Caltech, USA). ASM and BR would like to thank Inter-University Centre for Astronomy and Astrophysics (IUCAA) for their hospitality and facilities extended to them under their Visiting Associateship Programme.

\def\apj{ApJ}
\def\apjl{ApJl}
\def\pasp{PASP} \def\mnras{MNRAS} \def\aap{A\&A} \def\physerp{PhR} \def\apjs{ApJS} \def\pasa{PASA}
\def\pasj{PASJ} \def\nat{Nature} \def\memsai{MmSAI} \def\araa{ARAA} \def\iaucirc{IAUC} \def\aj{AJ} \def\aaps{A\&AS} 
\def\aapr{A\&ARv}
\bibliographystyle{mn2e}
\bibliography{aditya}

\end{document}